\documentclass[%
 reprint,
 amsmath,amssymb,
 aps,
]{revtex4-2}

\usepackage{amsmath}
\usepackage{graphicx}
\usepackage{dcolumn}
\usepackage{bm}
\usepackage[version=3]{mhchem}
\usepackage{mathtools}

\DeclarePairedDelimiterX\braket[2]{\langle}{\rangle}{#1 \delimsize\vert #2}
\usepackage{hyperref}
\usepackage{color,soul}

\begin{document}

\preprint{APS/123-QED}

\title{Thermodynamic Limits of Photon-Multiplier Luminescent Solar Concentrators}

\author{Tomi K. Baikie}
\author{Arjun Ashoka}
\author{Akshay Rao}
\author{Neil C. Greenham}
 \email{ncg11@cam.ac.uk}
\affiliation{%
 Cavendish Laboratory, University of Cambridge, Cambridge, CB3 0HE, UK
}%

\date{\today}

\begin{abstract}
Luminescent solar concentrators (LSCs) are theoretically able to concentrate both direct and diffuse solar radiation with extremely high efficiencies. Photon-multiplier luminescent solar concentrators (PM-LSCs) contain chromophores which exceed 100\% photoluminescence quantum efficiency. PM-LSCs have recently been experimentally demonstrated and hold promise to outcompete traditional LSCs. However, we find that the thermodynamic limits of PM-LSCs are different and are sometimes more extreme relative to traditional LSCs. As might be expected, to achieve very high concentration factors a PM-LSC design must also include a free energy change, analogous to the Stokes shift in traditional LSCs. Notably, unlike LSCs, the maximum concentration ratio of a PM-LSC is dependent on brightness of the incident photon field. For some brightnesses, but equivalent energy loss, the PM-LSC has a greater maximum concentration factor than that of the traditional LSC. We find that the thermodynamic requirements to achieve highly concentrating PM-LSCs differ from traditional LSCs. The new model gives insight into the limits of concentration of PM-LSCs and may be used to extract design rules for further PM-LSC design.
\end{abstract}

\maketitle

\section{Introduction}

\begin{figure}[t]
\includegraphics[width=0.9\linewidth]{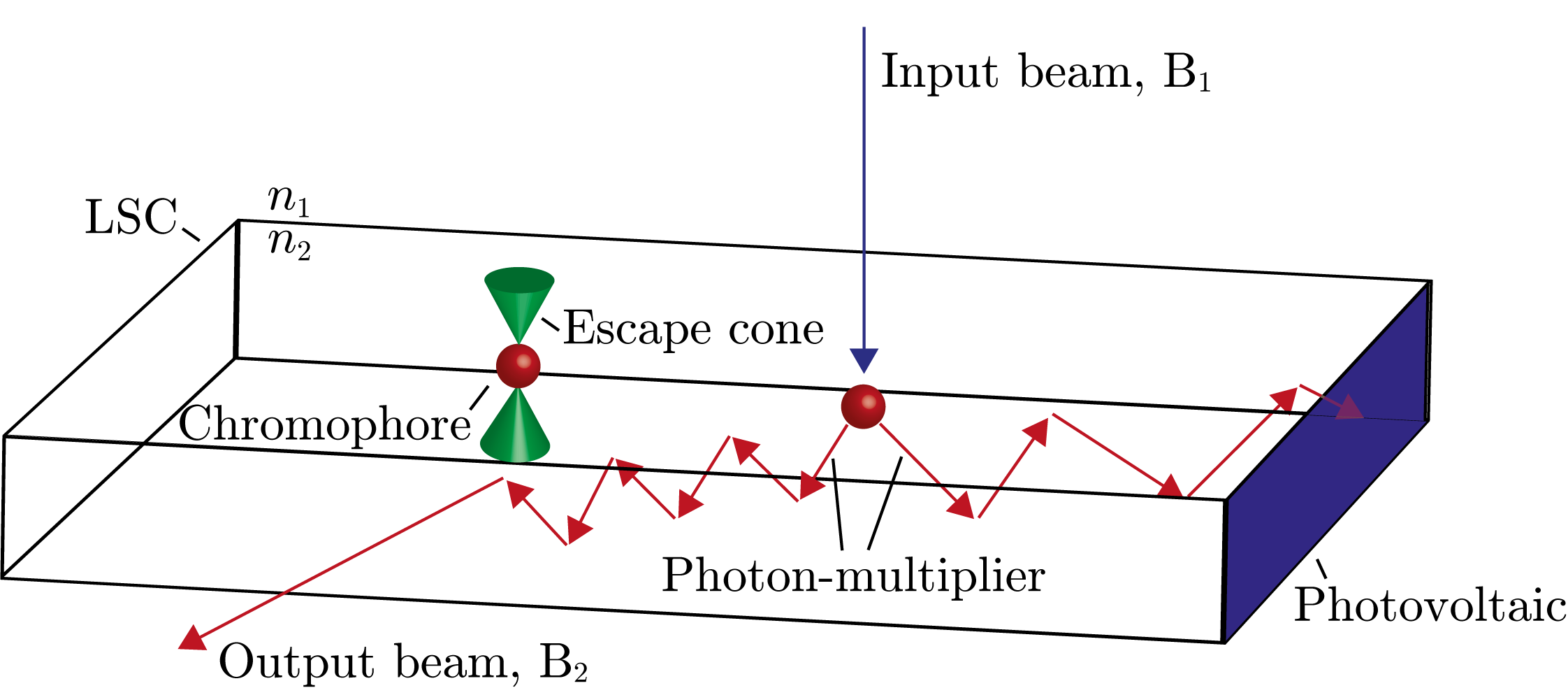}
\caption{\label{fig:epsart} \textbf{Schematic of operation of a photon-multiplier luminescent solar concentrators (PM-LSCs)}. As an input beam, $B_{1}$, irradiates the device, a chromophore absorbs and emits light at a longer wavelength. In a PM-LSC, maximally two photons are emitted for every incoming photon. Supposing isotropic emission, the number of photons lost through the escape cone is dependent on the ratio of the refractive indices of the outside medium ($n_{1}$) to the medium of the LSC ($n_{2}$). The photons which are emitted within the LSC are described by the second concentrated beam, $B_{2}$, which in proper operation impinges on a photovoltaic mounted on the edge of the LSC.}\label{figure_schematica}
\end{figure}

There is currently much interest in improving the yield of solar energy capture technologies, by making better use of the incoming solar radiation. One way to improve the solar cell effectiveness is to concentrate incoming solar irradiation \cite{Yablonovitch1980ThermodynamicsConcentrator,Debije2012ThirtyEnvironment, Rau2014ThermodynamicsDevices}. But in many regions of the world, the majority of the incident light is diffuse, but diffuse light cannot be concentrated by traditional image-preserving optical techniques, due to the limits established by the second law of thermodynamics \cite{Goetzberger1978FluorescentLight,vanHeerwaarden2021RecordWeather}. Traditional luminescent solar concentrators (LSC) offer an elegant way around this problem. An LSC is typically constructed from plastic or glass sheets with embedded chromophores, organic dyes or semiconductor nanocrystals, which efficiently absorb and emit light (Fig \ref{figure_schematica}). There is typically a free energy change which offsets the entropic loss when concentrating diffuse light. The LSC can thus concentrate both direct and diffuse light.

In traditional LSCs, a high-energy absorbed photon is converted to a lower-energy photon by the absorption and emission of a chromophore, inducing thermal loss in the form of a Stokes shift. The thermodynamics and efficiency limits of these systems have been previously described in detail \cite{Yablonovitch1980ThermodynamicsConcentrator,Papakonstantinou2015FundamentalYield,Batchelder1981LuminescentEfficiencies}. Recently, there have been proposals and experimental demonstrations of a new class of photon-multiplier luminescent solar concentrators (PM-LSCs), where high-energy photons are converted to two low-energy photons via processes such as quantum cutting in lanthanide-doped nanocrystals. Chromophores exhibiting up to 180\% PLQE have been experimentally demonstrated \cite{Luo2019Quantum-CuttingNanocrystals,Cohen2019Quantum-cuttingConcentrators}. Such systems could enable photoluminescence quantum efficiencies (PLQE) as high as 200\%.  The low reabsorption and high PLQE could make such systems ideal candidates for LSCs. There is currently no theoretical treatment of the thermodynamic limits of such a system, which are often useful to benchmark real device efficiencies against \cite{Shockley1961}.

Here, we present a thermodynamic model which gives insight into the fundamental limits of PM-LSCs. The maximum concentration performance of PM-LSCs is linked to brightness of the incoming photon field, which is not the case for traditional LSCs. Further, the maximum concentration factors associated with PM-LSCs are different, and at high brightness are more extreme, relative to their traditional counterparts.

\section{\label{sec:level2}Etendue and Entropy}

A beam of radiation, passing through an absorbing and luminescent medium, can be characterized by a temperature, chemical potential and etendue \cite{Landau1980StatisticalPhysics, Markvart2008TheEtendue}.  For a beam with ray direction within solid angle $\textrm{d} \Omega$ passing through an area $\textrm{d} A$, the element of etendue $\textrm{d} \xi$ is given by

\begin{equation}\label{eq:etendue}
	\textrm{d} \xi  = n^{2}\cos(\theta) \textrm{d} \Omega \textrm{d} A,
\end{equation}

\noindent where $\theta$ is the angle between the direction of propagation and the normal to $\textrm{d} A$, and $n$ is the refractive index of the medium. The etendue of a beam propagating through a clear and transparent medium is conserved. This conservation of etendue can be interpreted as the Liouville's theorem of classical mechanics, applied to photon propagation along rays in geometrical optics. Should absorption and remission occur, the etendue change is then a measure of the free energy loss. At moderate intensities of light, the thermodynamics of the system can be described by a two-dimensional gas where the number of dimensions arises from the two angular coordinates required to specify the direction of the beam \cite{Markvart2008SolarConversion}.

Notably, if entropy and etendue are constant, any concentration system which decreases the illuminated area $A$ must increase the angular divergence of the beam. In the case of diffuse light, the angular divergence is already maximal, and thus diffuse radiation cannot be concentrated with a system using only geometrical optics. In a traditional LSC, the change in enthalpy from heat dissipation associated with the Stokes shift can offset the entropic loss, allowing for the concentration of diffuse light with an associated energetic penalty.

The Heisenberg uncertainty principle, considered in the absence of diffraction, places a limit on the spread of the beam wavevectors in momentum space ; $\textrm{d}x \textrm{d}p_{x}\ge h$\cite{Markvart2008TheEtendue}. This condition naturally requires that only a single quantum state occupies the phase-space volume $\textrm{d}x \ \textrm{d}y \ \textrm{d}z \ \textrm{d}p_{x} \ \textrm{d}p_{y}\ \textrm{d}p_{z}$. A suitable choice of coordinate system gives the beam area, $\textrm{d}A$, as $\textrm{d}x\ \textrm{d}y$. The etendue element can then be written as

\begin{equation}
	\textrm{d} \xi = \frac{n^{2}}{k^{2}} \ \textrm{d}x\ \textrm{d}y\ \textrm{d}k_{x}\ \textrm{d}k_{y}.
\end{equation}

\noindent The volume element $\textrm{d}x\ \textrm{d}y\ \textrm{d}k_{x}\ \textrm{d} k_{y}$ in the phase space of variables $x,y,k_{x},k_{y}$ is equal to $ k^{2} \textrm{d} \xi/n^{2}$, where $n$ is the refractive index, and contains a single quantum state which can be occupied by a photon in a beam. For a system of $N$ photons, and allowing for two directions of polarization, we define

\begin{equation}
	G=\frac{2k^{2}}{n^{2}} \xi = \frac{2\nu^{2}}{c^{2}}\xi,
\end{equation}

\noindent which can be interpreted as the number of quantum states within a beam with frequency $\nu=ck/n$ and etendue $\xi$. These $N$ indistinguishable photons are distributed over the $G$ states according to,

\begin{equation}
W = \frac{(G+N)!}{G!N!} \approx \frac{(G+N)^{G+N}}{G^{G}N^{N}},
\end{equation}

\noindent where the Stirling's approximation has been used to obtain the second result. Invoking the ergodic hypothesis that each microstate is equally probable, the entropy of the beam, $S$, is the natural logarithm of the number of microstates, $W$, multiplied by the Boltzmann constant $k_{B}$,

\begin{equation} \label{eq:entropy_3_new}
\begin{split}
	S &= k_{B} \ln(W) \\
	&= k_{B} \frac{2 \nu^{2}}{c^{2}} \xi \Bigg(\bigg(1+ \frac{N}{G}\bigg)\ln\bigg(1+ \frac{N}{G}\bigg)-\frac{N}{G}\ln \bigg(\frac{N}{G}\bigg)\Bigg). \\
	\end{split}
\end{equation}

Following Yablonovitch's treatment, the entropy per photon, $s$, is the number differential of the entropy of the beam \cite{Yablonovitch1980ThermodynamicsConcentrator},

\begin{equation}\label{eq:entropy_change1}
      \begin{split}
	s=\frac{\textrm{d} S}{\textrm{d} N} &=
	k_{B}\ln\bigg(1+\frac{2\nu^{2}}{c^{2}} \frac{\xi}{N}\bigg).
	\end{split}
\end{equation}

 In the limit of maximum efficiency, where the loss of one photon from the incoming beam immediately results in the addition of two photons to the concentrated beam, the change in entropy of the concentrated photon beam then becomes

\begin{equation}\label{eq:entropy_change}
	\Delta S = 2 k_{B}\ln\bigg(1+\frac{2\nu^{2}}{c^{2}} \frac{\xi}{N}\bigg).
\end{equation}

\noindent Using the thermodynamic definition of entropy, $\delta Q/T$, rather than the statistical argument developed here, Equation \ref{eq:entropy_change} can be equivalently recovered by assuming that the photon gas is an effective heat bath and $\delta Q$ is the addition of 2 photons of energy $h\nu$.

\section{Brightness} \label{sec:brightness}

Radiation will come into thermal equilibrium with the chromophore provided that there is fast thermal equilibration among excited states   \cite{Yablonovitch1980ThermodynamicsConcentrator,Markvart2008TheEtendue}. As there are no first-order photon-photon interactions, photon beams cannot be in a true thermal equilibrium. Nevertheless, photon beams are well defined by thermodynamic parameters such as temperature, $T$, and chemical potential, $\mu$ \cite{Landau1980StatisticalPhysics}. The radiation emitted by a black body has zero chemical potential, however, quasi-blackbody radiation, where radiation is emitted over a restricted range, may have a non-zero chemical potential.  Planck derives the intensity of a monochromatic beam as

\begin{equation}\label{eq:intensity}
I = \frac{ h \nu^{3} n^{2}}{c^{2}} \frac{1}{e^{\frac{\mu - h\nu}{k_{B}T}}-1},
\end{equation}


\noindent with has units of energy per unit area, where $k_{B}$ is the Boltzmann constant and $T$ is the temperature of the medium in which emission from the chromophore takes place \cite[Eq. 300]{PlanckMax1991Ttoh}. As radiation in thermodynamic equilibrium is isotropic and unpolarised we can consider average chromophore absorption over $4\pi$ solid angle. Dividing Equation \ref{eq:intensity} by the energy of the photon ($h\nu$), we determine the photon flux of a beam as

\begin{equation}\label{eq:brightness_derv}
    B=\frac{8 \pi n^{2} \nu^{2}}{c^{2}} \frac{1}{e^{\frac{\mu - h\nu}{k_{B}T}}-1}.
\end{equation}

\noindent Equation \ref{eq:brightness_derv} is often referred to as the brightness of the beam and has units of photons per second, per unit bandwidth, per 4$\pi$ solid angle, per unit area \cite{PlanckMax1991Ttoh, Yablonovitch1980ThermodynamicsConcentrator, Landau1980StatisticalPhysics, Ross1967SomeSystems}. For monochromatic brightness we can write $\frac{\xi}{N}= \frac{4\pi n^{2}}{B}$. Where the input beam has some frequency range, the treatment presented by Markvart may be used \cite{Markvart2008TheEtendue}.  Brightness, like etendue, is conserved along the path of the beam in a perfectly transparent material. The transformation of one quasi-equilibrium beam into another introduces irreversibility, bringing about entropy generation.

\section{Concentration Limit}

The gain in entropy of the concentrated beam must be acquired from the free energy change associated with the photon-multiplicative process. For a traditional LSC with a chromophore of unity photoluminescence quantum efficiency, the concentration factor, C, is defined as the ratio of the brightnesses of the outgoing field, $B_{2}$, and the incoming field, $B_{1}$,

\begin{equation} \label{eq:trad_conc}
	C \equiv \frac{B_{2}}{B_{1}} \leq \frac{\nu_{2}^{2}}{\nu_{1}^{2}} \exp\bigg(\frac{h\big(\nu_{1}-\nu_{2}\big)}{k_{B}T}\bigg),
\end{equation}

 \noindent where $\nu_{1}$ and $\nu_{2}$ are the frequencies of the input photon and output photon, respectively \cite{Yablonovitch1980ThermodynamicsConcentrator}. The concentration limit of the PM-LSC can be determined from the second law of thermodynamics. Including the entropy source from the thermal loss, $h(\nu_{1}- 2\nu_{2})$,  the second law can be written

\begin{equation} \label{eq:reformed_second_law}
	\underbrace{2k_{B} \ln \Bigg(1+\alpha \frac{\nu_{2}^{2}}{B_{2}}\Bigg)}_{\Delta S_{2}}  + \frac{h(\nu_{1}-2\nu_{2})}{T} - \underbrace{k_{B} \ln\Bigg(1+\alpha \frac{\nu_{1}^{2}}{B_{1}}\Bigg)}_{ \Delta S_{1}} \geq 0,
\end{equation}

\noindent  where $\alpha = 8\pi n^2 /c^2$ and $\Delta S_{1}$ and  $\Delta S_{2}$ correspond to the entropy changes in the incident photon field and the generated photon field, respectively. Assuming the LSC is operating in a normal terrestrial environment, where $\alpha \nu^{2} / B \gg 1$, the ones in the logarithms can be ignored, and a concentration factor for PM-LSCs can be determined by rearranging Equation \ref{eq:reformed_second_law};

\begin{equation}\label{eq:pmf_conc_pre}
\begin{split}
  \frac{B_{2}^{2}}{B_{1}}&\leq \frac{8n^{2}\pi}{c^{2}}\frac{\nu_{2}^{4}}{\nu_{1}^{2}} \exp\bigg(\frac{h(\nu_{1}-2\nu_{2})}{k_{B}T}\bigg). \\
  \end{split}
\end{equation}

\noindent Hence the concentration factor is given by

\begin{equation}\label{eq:pmf_conc}
\begin{split}
  \frac{B_{2}}{B_{1}}&\leq \sqrt{\frac{8n^{2}\pi}{B_{1} c^{2}}\frac{\nu_{2}^{4}}{\nu_{1}^{2}} \exp\bigg(\frac{h(\nu_{1}-2\nu_{2})}{k_{B}T}\bigg)}. \\
  \end{split}
\end{equation}

\begin{figure}[t]
\includegraphics[width=0.9\linewidth]{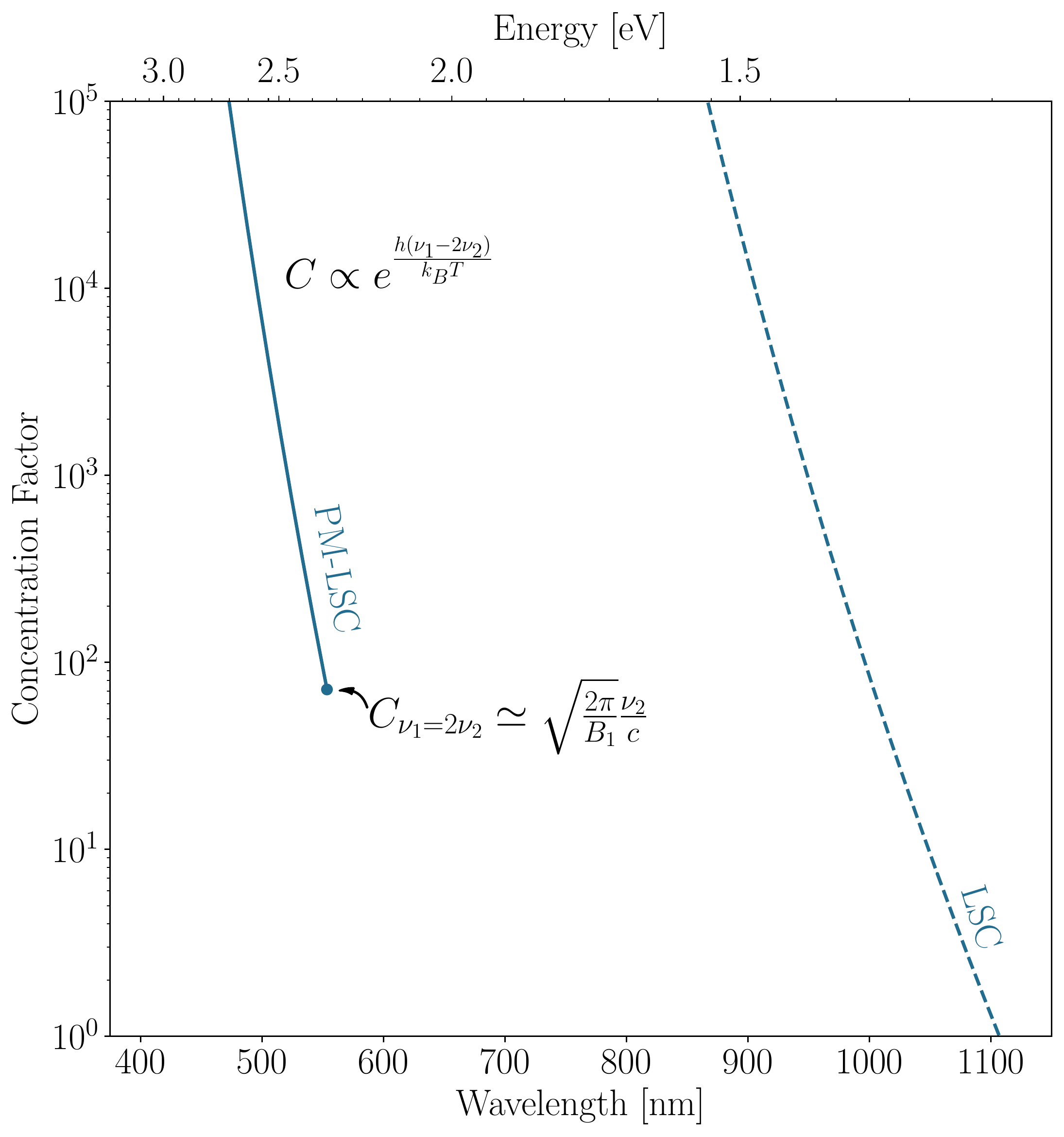}
\caption{ \textbf{Maximum concentration factor as function of chromophore absorption operating under terrestrial conditions}. The dotted and solid lines gives the maximum concentration factor, $C$, of the  traditional LSC and  PM-LSC, respectively, at room temperature ($T=300$ K) with $n = 1$, when illuminated with  light of brightness $10^{9}$ photons per second per unit bandwidth per unit area per $4\pi$ solid angle. This brightness roughly approximates a broad chromophore absorption over the terrestrial solar spectrum (see Appendix A). Emission for both the PM-LSC and the traditional LSC are fixed by $h \nu_{2} = 1.12 $ eV, to match the bandgap of silicon photovoltaics. In the PM-LSC, the onset of the photon-multiplication process occurs at $\nu_{1} = 2\nu_{2}$. In normal terrestrial conditions, the PM-LSC may concentrate at $\nu_{1} = 2\nu_{2}$ since the entropic change associated with the removal and addition of photons at different wavelengths ($\Delta S_{2}- \Delta S_{1}$) is greater than zero. When absorbing higher energy photons, $\nu_{1} > 2\nu_{2}$, the maximum concentration factor increases as the thermal loss is increases, analogous to the Stokes shift in the traditional LSC.}\label{figure1_new}
\end{figure}

Surprisingly, we find that for PM-LSCs the concentration limit is dependent on input brightness. Comparing the concentration limit of a PM-LSC, Equation \ref{eq:pmf_conc}, to that of a traditional LSC, Equation \ref{eq:trad_conc}, we find similar design requirements, but also some important distinctions \cite{Yablonovitch1980ThermodynamicsConcentrator}.

In a traditional LSC, when the Stokes shift is zero ($\nu_{1} = \nu_{2}$), there will be no associated thermal loss as $h(\nu_{1}-\nu_{2})/T =0$, and the difference in the entropic terms associated with the addition and removal of the photon will equal, $\Delta S_{2}- \Delta S_{1}=0$, thus the LSC is unable to concentrate. However, in the PM-LSC, particularly at low brightnesses, may concentrate, even if there is no thermal loss, i.e. $h(\nu_{1}-2\nu_{2})/T =0$. Even at the energy conservation limit, $\nu_{1} = 2 \nu_{2}$, there remains a source of entropy from the addition and removal of photons of different wavelengths, $\Delta S_{2}- \Delta S_{1}$) which is sufficient to allow for concentration.

To increase the concentration limit of a PM-LSC, or to enable effective concentration at higher brightnesses, an additional thermal loss is required to satisfy the inequality in Equation \ref{eq:pmf_conc}, analogous to the Stokes shift in the traditional LSC. As depicted in Figure \ref{figure1_new}, the concentration factor is exponentially dependent on the thermal loss in PM-LSCs. Figure \ref{figure1_new} assumes that both the PM-LSC and LSC chromophore emission is given by $h\nu_{2} = 1.12$ eV, and thus optimised for the silicon photovoltaic bandgap ($E_{g} = 1.12$ eV). Both the traditional LSC and the PM-LSC are assumed to have the same matrix medium with a refractive index of $n=1$. In the case of PM-LSCs, only photons of energy of $h\nu_{1}\ge 2.48$ eV may undergo a PM processes to comply with the constraints imposed by the conservation of energy. The input brightness, $B_{1}$, is fixed at $10^{9}$ photons per second per unit bandwidth per unit area per $4 \pi$ solid angle. This brightness represents a broad absorption over the terrestrial solar spectrum for the PM-LSC, though there exists difficulties in determining this value when approximating solar radiation over some narrow bandwidth (see Appendix A). Care must be taken when modelling much greater brightnesses, for example where the input solar flux has been concentrated before entry into the LSC, or for extremely small bandgap materials, where Equation \ref{eq:pmf_conc} must be reformed, as $\alpha \nu^{2} / B \gg 1$ is no longer valid (see Appendix B).

We have identified that to achieve very high concentration factors, as for traditional LSCs, a PM-LSC design may also include a thermal loss. Nevertheless, for some brightnesses, but equivalent thermal loss, the PM-LSC has a greater maximum concentration factor than that of the traditional LSC.

\begin{figure}[t]
\includegraphics[width=0.9\linewidth]{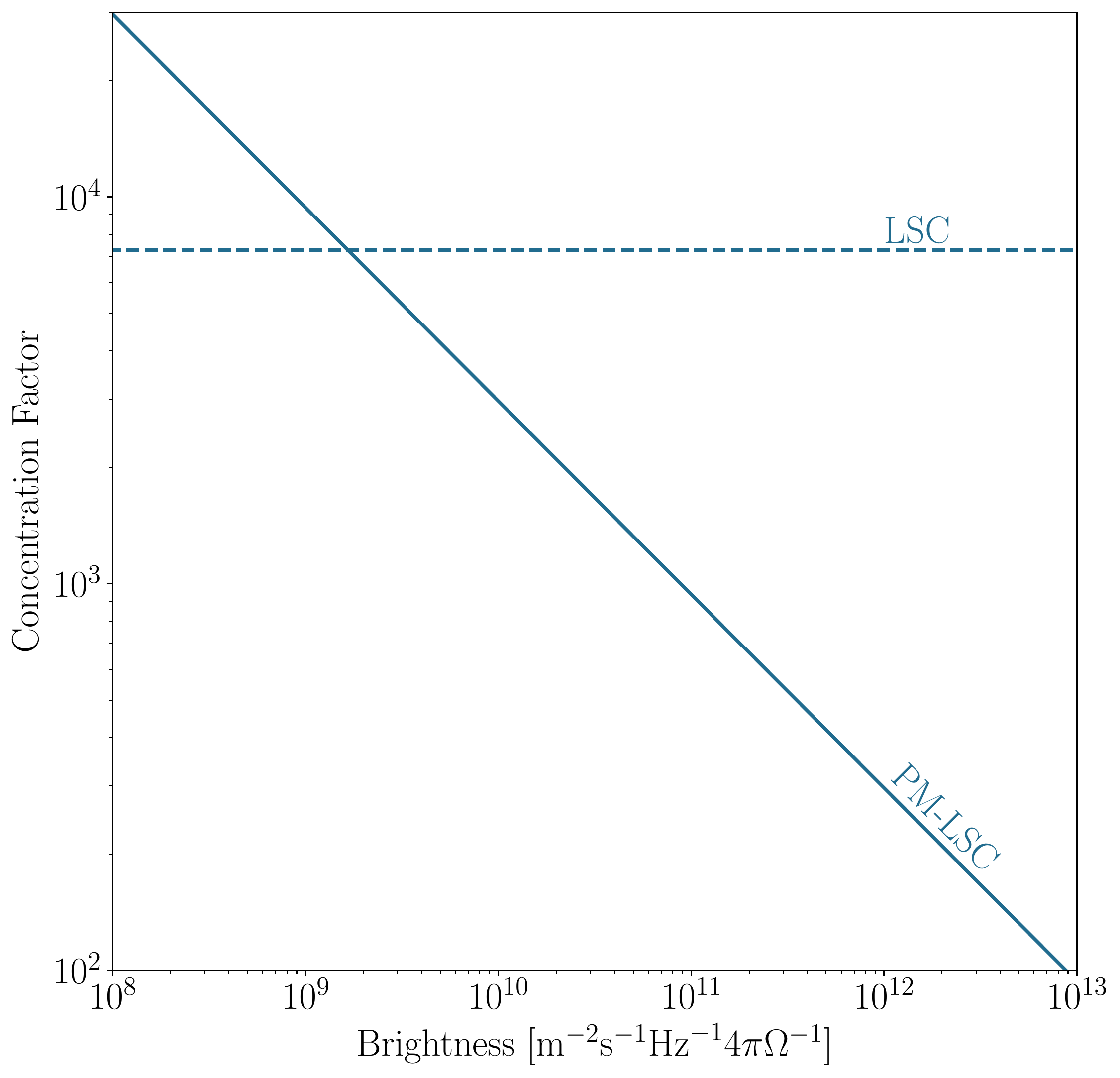}
\caption{\textbf{Maximum concentration factor at room temperature for  a  traditional  LSC  (dotted)  and  a  PM-LSC  (solid)  as function of incident  brightness}. Both the PM-LSC and the traditional LSC concentration limit have emission at $h\nu_{2} = 1.12$ eV, optimised for silicon photovoltaics. The traditional LSC has a chromophore with a Stokes shift of $0.24$ eV ($ \lambda_{1} = 912$ nm) which gives a brightness independent maximum concentration factor $C=7280$. The PM-LSC is assumed to be 200 percent efficient. The PM-LSC has a thermal loss matching that of the traditional LSC, i.e. $h\nu_{1} = 2 \times 1.12 + 0.24 = 2.48$ eV input, corresponding to monochromatic incident light of wavelength $\lambda_{1} = 500$ nm. }
\label{brightnessdependence2}
\end{figure}

Figure \ref{brightnessdependence2} outlines the effect of input brightness on concentration factor. Here again, the emission is optimised for a silicon photovoltaic bandgap ($h\nu_{2} = 1.12$ eV). In this case, however, the thermal loss is fixed at $0.24$ eV for both the traditional LSC and the PM-LSC. The fixed thermal loss defines the monochromatic photon absorption at $\lambda_{1} = 912$ nm in the traditional LSC and at $\lambda_{1} = 500$ nm in the PM-LSC. For this specific thermal loss, only weak brightnesses allow for a maximum PM-LSC concentration factor greater than the traditional LSCs. Although the maximum concentration factor for PM-LSCs may increase more rapidly with increasing thermal loss relative to the traditional LSC, at very high brightness the PM-LSC cannot reach the same maximum concentration ratios as that of the traditional LSC.

\begin{figure}[t]
\includegraphics[width=0.9\linewidth]{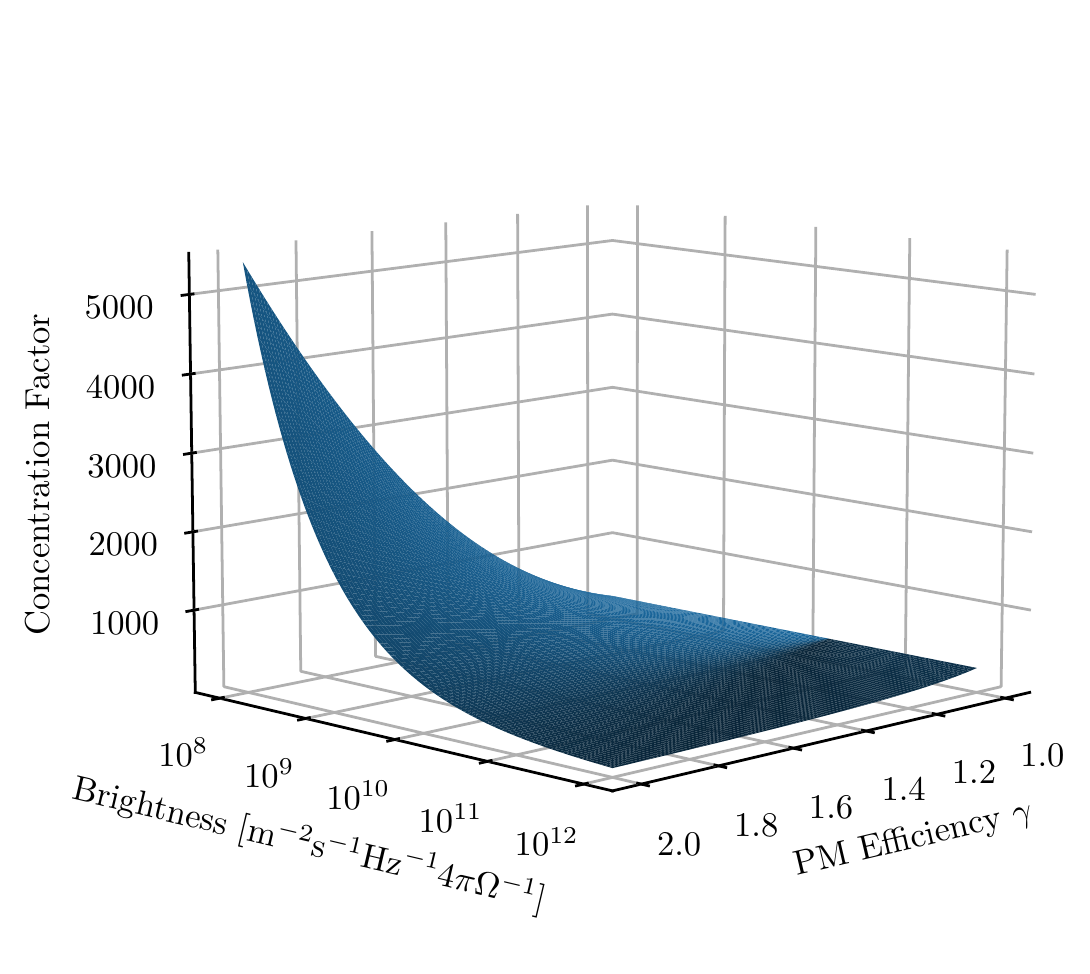}
\caption{\textbf{Maximum concentration as function of PM efficiency and input brightness}. The blue surface represents the maximum concentration factor for a PM-LSC with a variable PM efficiency, $\gamma$, under different incident brightness. $\gamma$ varies from 1, which would be the maximum efficiency of a traditional LSC, to 2, the maximum efficiency of PM-LSCs considered here. The emission frequency for the PM-LSC is fixed at $h\nu_{2} = 1.12$ eV, to match that of silicon photovoltaics. Incident illumination is described by a monochromatic beam with wavelength $h\nu_{1} = 1.12\gamma + 0.15$ eV. The input brightness is plotted between $10^{8}$ to $10^{12}$ photons per second, per unit bandwidth, per unit area, per $4\pi$ solid angle.}
\label{map}
\end{figure}

Significant research effort has been expended to improve the efficiencies and understand the mechanisms of photon-multiplier technologies \cite{Rao2017, Milstein2018PicosecondNanocrystals}. As such, it is of general interest to determine how material design improvements may impact on LSC uses and design. Figure \ref{map} plots PM-LSC concentration as a function of $\gamma$, which is a measure of the PM-system efficiency between $1 \le \gamma \le 2$ (see Appendix C for full expression).  A maximally efficient traditional LSC chromophore will have $\gamma=1$, meaning one photon absorbed will result in one photon emitted by the chromophore, whereas a PM-LSC with $200\%$ photon conversion efficiency has $\gamma=2$. We note that while the notion of a fractional $\gamma$ is not physical in the single photon limit, in the thermodynamic limit which we study, it is purely a ratio of underlying photon-yielding processes. For example, in organic semiconductors, the interplay between singlet fission or inter-system crossing pathways for photoexcited singlets sets the effective $\gamma$ of the system, with pure singlet fission yielding $\gamma = 2$ and pure inter-system crossing yielding $\gamma = 1$, assuming the triplets eventually yield luminescent species of interest\cite{Rao2017}.

At low brightness, even a small improvement in $\gamma$, from $1$ to $1.2$ would offer a two-fold increase in the maximum concentration factor of PM-LSCs, implying that moving away from traditional LSCs to even weakly efficient PM process yields gains in the concentration factor. For the highest brightnesses, however, a diminishing return on the investment of improving $\gamma$ is made. Taking the example of singlet fission photon-multiplier systems, this suggests that even a small improvement in the singlet fission process at low brightnesses could lead to impressive improvements in the maximum concentration factors of PM-LSCs. This also suggests PM-LCSs may play an important role in low brightness energy harvesting systems.

Further, this work suggests that exothermic singlet fission chromophores may lead to higher maximum concentration factors relative to endothermic singlet fission chromophores \cite{Rao2017}. Without the inclusion of another heat source, endothermic singlet fission chromophores would act to reduce the entropic loss in Equation \ref{eq:reformed_second_law} and hence reduce the concentration limit.

Practical demonstrations of PM-LSCs utilise quantum cutting, where ytterbium doped perovskite nanocrystals feature a photoluminescence quantum yield approaching 200\% and virtually zero self-absorption loss \cite{Luo2019Quantum-CuttingNanocrystals,Cohen2019Quantum-cuttingConcentrators}. These systems show great promise for PM-LSCs, however, the ytterbium-doped perovskite chromophores exhibit long-lived excited states on the order of milliseconds. These systems violate our assumption that the chromophore exhibits quick electronic recovery to a photon accepting state \cite{Erickson2019PhotoluminescenceDownconversion}. While these systems exhibit long lived excited states, they will not reach the concentration limits described here.

In practice, both PM-LSCS and traditional LSCs are far from being limited by the entropic arguments put forward here. Record demonstrations of traditional LSCs achieve concentration ratios on the order of 1-10 \cite{Roncali2020LuminescentVadis}. Although at high brightness the PM-LSC concentration limit is significantly below that of traditional LSCs, it is still on the order of $10^{2}-10^{3}$. Further, without utilisation of hot carrier solar cells coupled to LSCs, Auger recombination in standard photovoltaics will play a significant role at high photon flux, potentially hampering the positive gain of large concentration factors \cite{Nelson2003TheCells}.

\section{Conclusion}
\noindent  Considering the propagation and transformation of radiation in the thermodynamic limit, we have obtained the entropy generation rate for a photon-multiplicative process. From the second law of thermodynamics we have obtained the concentration limit for an ideal PM-LSC. Unlike traditional LSCs, the maximum concentration in a PM-LSC is dependent on brightness of the incident field. For very high concentration factors, particularly in bright conditions, the PM-LSC will require a large thermal loss, analogous to the Stokes shift in a traditional LSC. We find at low brightness, the PM-LSC may exceed the concentration limit of traditional LSCs. Further, at low brightness, the improvements required in the photon multiplier efficiency need only be small to have a dramatic impact on the concentration limit.

\vspace{19mm}
\begin{acknowledgments}
The authors are thankful to the Engineering and Physical Sciences Research Council (EPSRC) for financial support. T.K.B. recognises the support of the Centre for Doctoral Training in New and Sustainable Photovoltaics. A.A acknowledges funding from the Gates Cambridge Trust and the as well as support from the Winton Programme for the Physics of Sustainability.
\end{acknowledgments}
\vspace{3mm}

\bibliographystyle{unsrt}

\end{document}